\documentclass[12pt]{iopart}
\usepackage[latin1]{inputenc}
\usepackage{graphicx}
\usepackage{dcolumn}
\usepackage{bm}

\begin{document}
\title["Magnetoscan": A Modified Hall Probe Scanning Technique]{"Magnetoscan": A Modified Hall Probe Scanning Technique for the Detection of Inhomogeneities in Bulk High Temperature Superconductors}
\author{M.~Eisterer, S. Haindl, T. Wojcik and H.~W.~Weber}
\address{Atominstitut der Österreichischen Universitäten, A-1020
Vienna, Austria}

\ead{eisterer@ati.ac.at}

\begin{abstract}
We present a novel technique for the investigation of local
variations of the critical current density in large bulk
superconductors. In contrast to the usual Hall probe scanning
technique, the sample is not magnetized as a whole before the
scan, but locally by a small permanent magnet, which is fixed near
the Hall probe, during the scanning process. The resulting signal
can be interpreted as a qualitative measure of the local shielding
currents flowing at the surface.
\end{abstract}



\section{Introduction}
Melt textured bulk superconductors are promising candidates for
technical applications such as permanent magnets, motors or
flywheels. These applications require macroscopically homogeneous
properties in the largest possible specimens. However, melt
processed materials are strongly inhomogeneous, at least on a
microscopic scale \cite{Dik95}, since the formation of defects,
e.g. 211 particles, microcracks, twin planes, subgrain boundaries,
stacking faults, dislocations, oxygen deficient regions or the
boundaries between the growth sectors, cannot be avoided during
the growth process. These defects influence the transport
properties in various ways. Large normal conducting inclusions
simply reduce the superconducting cross section, cracks impede the
current flow locally, (sub)grain boundaries reduce the critical
current density across them \cite{San01}, but some defects can
also act as pinning centers and, therefore, improve the transport
properties. The defect structure is related to the growth
conditions \cite{Sal89,Oga91}. Differences were found between the
a-growth and the c-growth sectors \cite{Dik03} and the defect
concentration generally increases with increasing distance from
the seed.

The experimental characterization of the transport properties in
large samples is commonly made by scanning the remnant flux
profile. This technique is rather insensitive to material
inhomogeneities \cite{Gon03}, since all parts of the sample volume
contribute to the magnetic field at each measured point. Even if
the current distribution can be estimated under certain
assumptions \cite{Car03}, it is generally impossible to assess the
correlation between the critical currents and the local
microstructure, since $J_{c}$ does not only depend on the
microstructure, but also on the magnetic field, which changes
within the sample. Therefore, a new technique - the "magnetoscan"
- has been developed. Here, the currents are induced only locally
over a small sample area, where the magnetic field remains nearly
constant. Moving the magnet enables us to compare the local
transport properties at different positions of the sample with a
spatial resolution of less than one millimeter. The magnetoscan,
therefore, represents an ideal tool to relate the local $J_{c}$ to
microstructural investigations, in order to optimize the growth
process in the end.

\section{The principle: Local shielding currents induced by a permanent magnet}
Although the details of the response of a superconductor to a
nonuniform magnetic field induced by a permanent magnet are quite
complex \cite{Qin02}, shielding can be understood qualitatively,
if the permanent magnet is much smaller than the superconductor
and if the magnet is placed close to the surface, but not close to
the edge of the superconductor.

\begin{figure} \centering
\includegraphics[clip,width = \columnwidth]{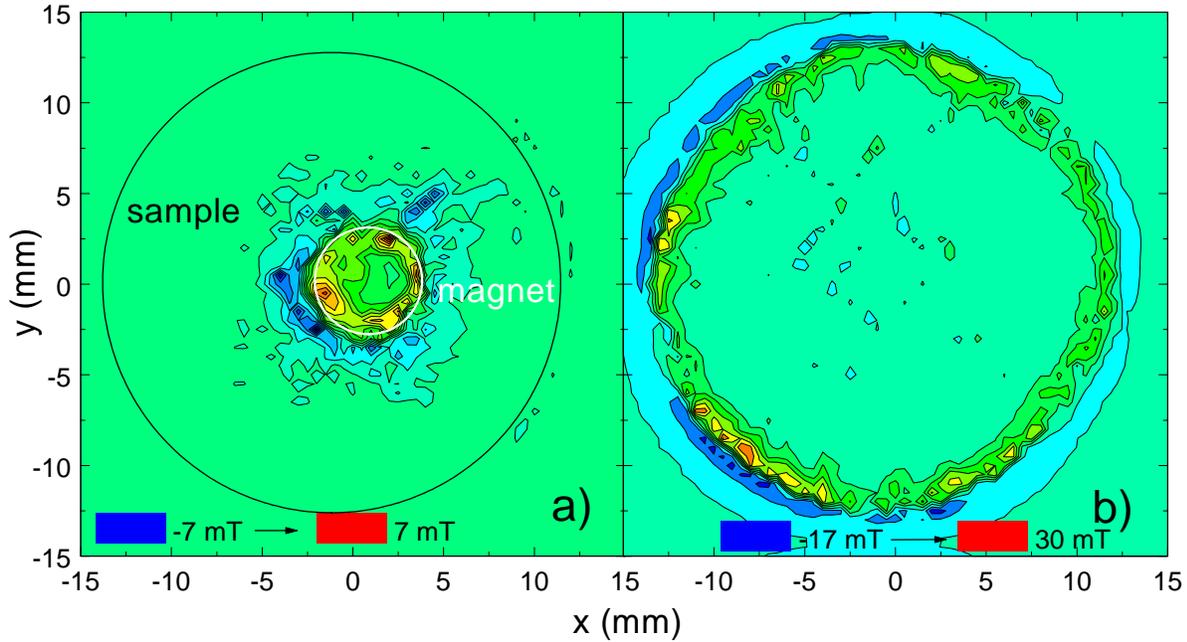} \caption{(a)
Trapped flux profile after touching the surface of the
superconductor with a permanent magnet. The white and black
circles indicate the position of the magnet and sample,
respectively. (b) Remnant flux after a magnetoscan recovered by
conventional field mapping.} \label{fig1}
\end{figure}

The following experiment was made. A cylindrically shaped
permanent magnet was moved vertically to the surface of a large
bulk sample ($R=12.9\,mm$, $h=10.8\,mm$), which had been zero
field cooled in liquid nitrogen. The magnet with a radius
$R=3\,mm$ and a height $h=18\,mm$ had a magnetic induction of
$100\,mT$ at the center of its top surface. After removing the
magnet, the remnant flux profile of the sample was scanned by a
conventional Hall probe scanner at a distance of $0.2\,mm$ above
the surface (Fig. \ref{fig1}a).
Flux penetrated the sample near the edge of the magnet (white
circle), whereas no flux is trapped in the center (the inner
profile is built up by removing the magnet). Flux penetrates and
leaves (bright spots) the sample very inhomogeneously, which must
be related to the local properties of the sample.

This observation represents the basis for the method presented in
this paper. Flux penetrates more easily in regions with "bad"
transport properties, the response of the sample is smaller there.
In regions with high shielding currents, the feedback field of the
sample is larger. The flux penetration at the edge of the sample
(black circle in Fig. \ref{fig1}a) is not important for our
purpose and results from the large demagnetization there. If the
magnet is moved horizontally over the sample surface, the flux
profile could either move simultaneously or flux could (partly) be
trapped in the superconductor. This can be checked by mapping the
remnant flux after moving the magnet parallel to the sample
surface along the typical path of a Hall scan (Fig. \ref{fig1}b).
We find that flux is trapped at the edge of the sample, but
(almost) not inside. This indicates that hysteretic memory effects
due to the scanning process are absent, apart from the sample
edge.

The penetration depth of the magnetic field is roughly estimated
from the relation $\delta =B/\mu_{0}J$. With a typical critcal
current density of $10^{8}\,Am^{-2}$, the penetration depth of the
field of this permanent magnet is expected to be slightly smaller
than $1\,mm$.

\section{Magnetoscan: Experimental Setup and Tests}

\begin{figure} \centering \includegraphics[clip,width
= 0.4\columnwidth]{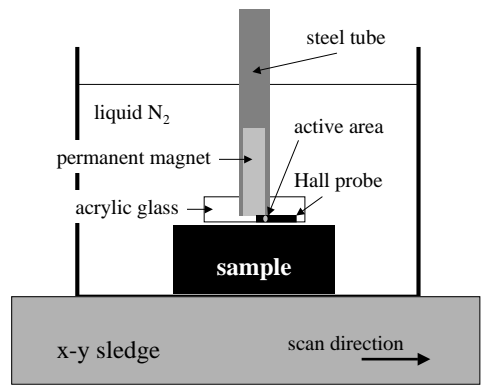} \caption{Modified Hall probe scanning
device} \label{fig2}
\end{figure}

Fig. \ref{fig2} shows the experimental setup schematically. A Hall
probe (AREPOC HHP-VPO) with an active area of $50\times 50\, \mu
m^{2}$ is fixed on the edge of the small permanent magnet used
above. The permanent magnet and the Hall probe remain fixed, while
the sample is moved together with the dewar by a x-y sledge. The
gap between the active area of the Hall probe and the sample
surface is about $0.2\,mm$, between the permanent magnet and the
sample about 1 $mm$. Measurements of the Hall voltage were taken
on a 0.25 or $0.5\,mm$ grid.
A preliminary scan was made with the Hall probe positioned at the
center of the magnet's surface, not at the edge. In this case the
signal nearly did not change apart from the sample edge. This
behavior is expected from the above considerations about the
magnet's "fingerprint", since the magnetic field was found to be
zero at the center of the magnet. Only the transition from nearly
perfect screening above the sample to no screening beside it is
observed.

Figure \ref{fig3}a shows a typical example of a magnetoscan (with
the Hall probe at the edge of the magnet). The arrow  indicates
the direction of the relative movement of the magnet and the Hall
probe. The Hall probe always follows the magnet (when measurements
are taken) in order to minimize the remaining hysteretic effects.
The change of the signal is much faster, when the Hall probe
reaches the sample, since the magnet is already above the sample
in this case, while on the opposite edge, the magnet has already
left the sample. Due to the rapidly changing geometrical
conditions, the signal cannot be evaluated in terms of current
transport properties there. A further magnetoscan was made after
rotating the sample by 180°. As expected, the geometrical effects
near the sample edge remained unchanged, whereas the details of
the magnetoscan caused by inhomogeneities were also rotated by
180°.

It should be mentioned that the signal is very sensitive to the
gap between the Hall probe and the sample surface. Only samples
with a flat surface can be investigated. If the (flat) sample
suface is not oriented completely horizontally, the signal
increases with decreasing gap, but the inhomogeneities are still
detectable, because the systematic influence of such a
misalignment and of local changes caused by inhomogeneities can be
easily distinguished.

\section{Results}

\begin{figure} \centering \includegraphics[clip,width
= \columnwidth]{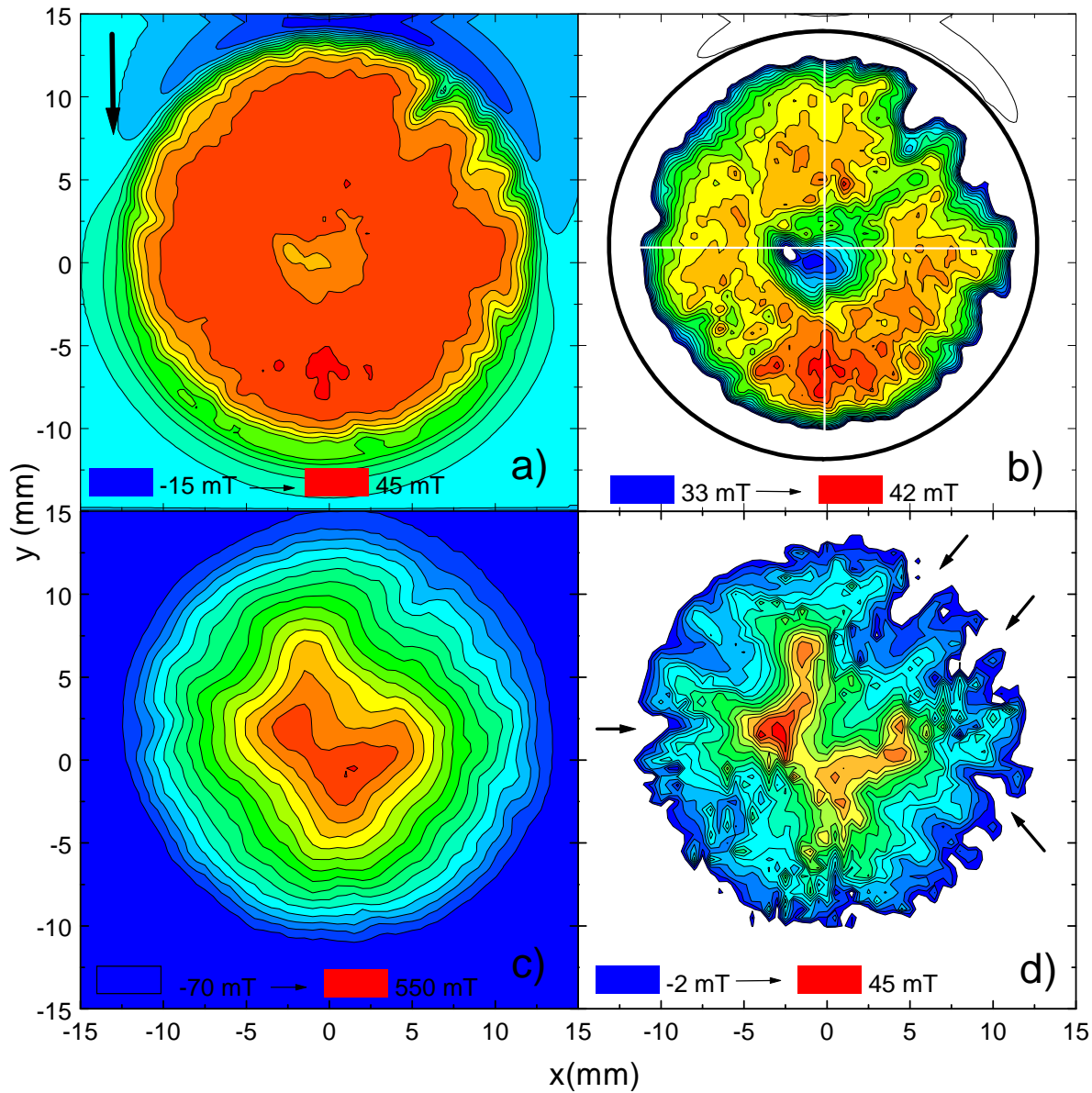} \caption{(a,b) Example of a magnetoscan,
plotted on different scales. The solid circle indicates the
circumference of the sample, the white lines indicate the
boundaries of the growth sectors. (c) Usual trapped flux profile
of the same sample. (d) Difference betweem the trapped flux
profiles measured at $0.2\, mm$ and $0.4\, mm$ above the sample
surface.} \label{fig3}
\end{figure}

Figure \ref{fig3}b shows the details of the magnetoscan discussed
above (Fig.\ref{fig3}a) just for the inner part of the sample. The
difference in the Hall voltage between two contour lines is now
smaller by a factor of 7 compared to Fig. \ref{fig3}a. The usual
trapped flux profile ($0.2 \, mm$ above the sample surface) is
plotted for comparison in Fig. \ref{fig3}c ($B_{T}^{max}=0.54 \,
T$).
Three interesting features can be observed in this magnetoscan:
the irregularly shaped contour lines, the smaller response at the
position, where the seed was located, and the highest values near
the boundaries between the different a-growth sectors (white lines
in Fig. \ref{fig3}b).

In order to confirm the relation of these findings to the local
transport properties, the conventional trapped field profile was
measured $0.2\, mm$ and $0.4\, mm$ above the sample surface. The
difference of these two profiles should be approximately
proportional to the derivative of $B_{z}$ with respect to the
vertical coordinate, i.e. $\partial B_{z}/
\partial z$. This quantity is strongly influenced by the surface
currents as follows from the Biot-Savart law:
\begin{equation} \nonumber
{\bf B}({\bf r})=\frac{\mu_{0}}{4\pi}\int\limits_{V}\frac{{\bf
J}({\bf r}')\times ({\bf r}-{\bf r}')}{|{\bf r}-{\bf r}'|^{3}}\,
d^{3}r'
\end{equation}
The integrand does not depend on $z$ alone, but always on $z-z'$.
Therefore, in the integral for ${\bf B}({\bf r}+\delta{\bf
e}_{z})$ (in the difference quotient) ${\bf r}$' can be
substituted by ${\bf r}'-\delta{\bf e}_{z}$, and, if $\partial
{\bf J}/\partial z$ exists, the partial derivative of $\bf B$ with
respect to $z$ can be expressed by
\begin{eqnarray}
\frac{\partial {\bf B}}{\partial z}({\bf
r})&=&\frac{\mu_{0}}{4\pi}\int\limits_{F}\frac{{\bf J}({\bf
r}')\times ({\bf r}-{\bf r}')}{|{\bf r}-{\bf r}'|^{3}}
\bigg{|}_{z'=-h/2}\, dx'dy'\\ \nonumber
&&-\frac{\mu_{0}}{4\pi}\int\limits_{F}\frac{{\bf J}({\bf
r}')\times ({\bf r}-{\bf r}')}{|{\bf r}-{\bf
r}'|^{3}}\bigg{|}_{z'=h/2}\, dx'dy'\\ \nonumber
&&+\frac{\mu_{0}}{4\pi}\int\limits_{V}\frac{\partial {\bf
J}}{\partial z}\,\bigg{|}_{{\bf r}={\bf r}'}\times\frac{({\bf
r}-{\bf r}')}{|{\bf r}-{\bf r}'|^{3}}\, d^{3}r'
\end{eqnarray}

The first and the second term are integrations over the bottom
($z'=-h/2$) and top ($z'=h/2$) surface of the sample,
respectively. If the field profiles are measured close to the top
surface, the first term can be neglected. The influence of the
third term is a priori unknown. If the variation of the currents
along the sample thickness is smooth (at least near the top
surface), which ensures the existence of $\partial {\bf
J}/\partial z$, its contribution is small, otherwise it could
become dominant. If it is negligible, the difference of two field
profiles measured at slightly different heights (Fig.
\ref{fig3}d), is proporional to the field profile of the surface
currents (second term). The qualitative agreement with the
magnetoscan is satisfactory. Note that the two profiles are not
expected to be identical, since currents all over the surface
contribute to the differential profile (plus a contribution of the
whole sample), while the currents are induced only locally during
the magnetoscan. For instance, a completely homogeneous current
density results in a constant signal during the magnetoscan (apart
from the edges), but the corresponding differential profile has a
sharp peak at the center of the sample (field of a thin sample).
Nevertheless inhomogeneities can be detected in both cases. The
regions with weak currents near the sample edge (indicated by
arrows in Fig. \ref{fig3}d), the higher currents at the boundaries
of the a-growth sectors and the reduction of the currents directly
below the seed are clearly observed with both methods.

\begin{figure} \centering \includegraphics[width
= \columnwidth]{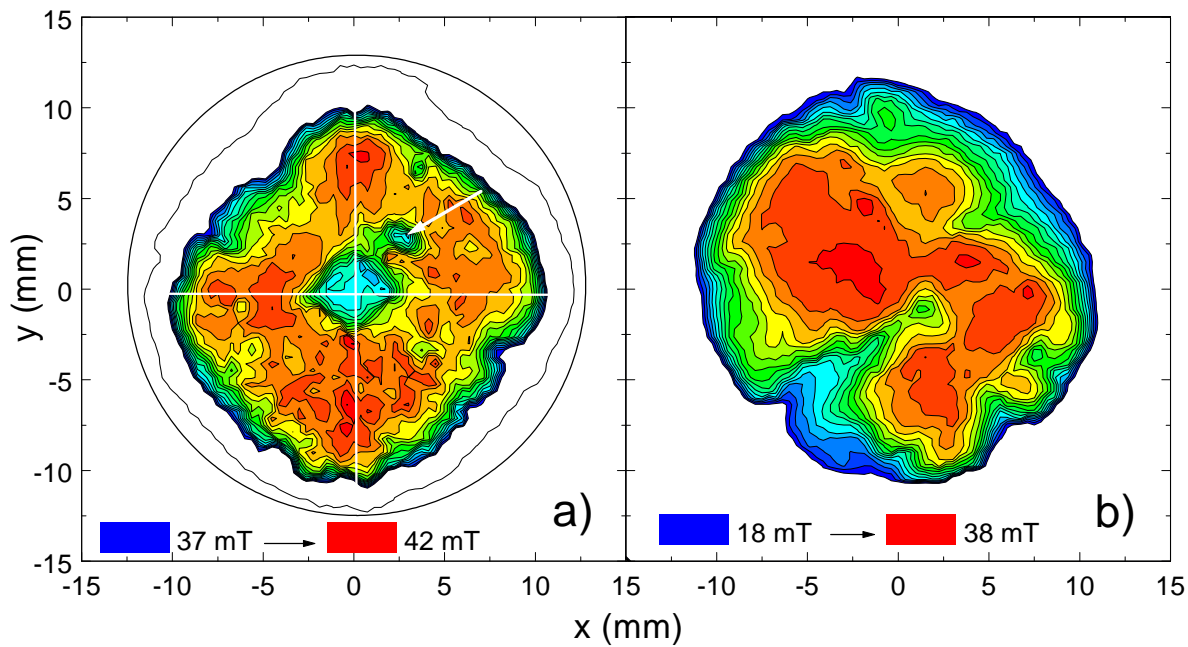} \caption{Magnetoscan (a) of the top
side, (b) of the bottom side of the sample.} \label{fig4}
\end{figure}

Further examples of magnetoscans are plotted in Fig. \ref{fig4}.
The dimension of the sample was $R=13.1\,mm$ and $h=9.8\,mm$, its
maximum trapped field after dc activation was $0.79\,T$.
The profile of the top side (a) is nearly quadratic, although the
sample is cylindrically shaped. The best shielding is again
observed near the growth boundaries (oriented horizontally and
vertically) and a depression of the shielding currents is found at
the former position of the seed. The weaker shielding near the
seed (indicated by the white arrow) corresponds to a bright spot,
which is visible at the sample surface. At the bottom side (Fig.
\ref{fig4}b) the transport properties are inhomogeneous on a
larger scale. By optical inspection of the sample, two types of
differently looking regions are observed at the bottom side, i.e.
crystalline regions with a pronounced subgrain structure and
rather dull black looking areas. The shape of these areas is
nicely displayed by the magnetoscan with higher shielding currents
in the crystalline regions.

Although many inhomogeneities can be ascribed to optically visible
features of the samples (position of the seed, growth sector
boundaries etc.), there remain peculiarities in the results, which
cannot be correlated with visible inhomogeneities, e.g. the
reasons for the irregular (cf. \ref{fig3}b) or quadratic (cf.
\ref{fig4}a) shape of the magnetoscan profiles.

\section{Conclusions}

Magnetoscanning represents a simple method to investigate changes
of the transport properties of large bulk superconductors with a
spatial resolution of around one millimeter. In contrast to the
usual scanning Hall probe method, shielding currents are induced
only locally. The range of magnetic field penetration can be
chosen by the size and the strength of the permanent magnet or
could even be varied by using an electromagnet.

\end{document}